# Tau lepton physics at LEP

**Zhiqing Zhang**[*]
*Laboratoire de l'Accélérateur Linéaire, IN2P3/CNRS et Université de Paris-Sud XI*
*BP34, F-91898 Orsay, France*
*E-mail:* `zhangzq@lal.in2p3.fr`

The talk covers three contributions on (i) the final measurement of branching ratios and spectral functions of $\tau$ decays using the full LEP-I data from ALEPH, (ii) a preliminary measurement of the $\tau$ hadronic branching ratios from DELPHI and (iii) a measurement of the strange spectral function in hadronic $\tau$ decays from OPAL. These measurements are discussed and the relevant physics topics are briefly reviewed.

---

[*] On behalf of the ALEPH, DELPHI and OPAL Collaborations

## 1. Introduction

The tau lepton, the heaviest lepton known so far, has been studied in great detail at LEP. Indeed the LEP-I machine running at center of mass energies around the $Z^0$ resonance provides a clean environment allowing high $\tau$ selection efficiency and low non-$\tau$ background contribution for measuring properties of the tau. Two examples are branching ratios (BRs) of both leptonic and hadronic decay modes and spectral functions (SFs or unfolded invariant mass distributions) of the hadronic final state. The corresponding measurements have been thoroughly exploited in testing universalities between $e$, $\mu$ and $\tau$ and providing the most precise measurement of fundamental parameters in QCD [1]. This talk describes three contributions received to this conference from ALEPH [2], DELPHI [3] and OPAL [4] and discusses some of the physics implications of the measurements [1].

## 2. Recent measurements at LEP

A complete and final analysis of $\tau$ decays has been published recently by ALEPH [2] which marks the end of an enormous effort during more than a decade in making best use of the detector capability. Based on the full LEP-I data taken during 1991-1995 and using a global method, all decay modes having a BR greater than 0.1% have been exclusively classified. In total 13 decay modes have been determined including $e$ and $\mu$ leptonic decay modes. The best precision reaches down to 0.4%. In addition, the SFs for 5 dominant nonstrange hadronic decay modes ($\pi\pi^0$, $3\pi$, $\pi2\pi^0$, $3\pi\pi^0$, $\pi3\pi^0$) have been measured. These measurements complete previously measured BRs [5, 6] and SFs [6] of strange decay modes.

A preliminary measurement of hadronic BRs has been obtained by DELPHI [3] for 6 exclusive and 6 semi-exclusive decay modes using a neural network analysis (cross-checked by a classical sequential cut analysis). The leptonic BRs were previously measured in [7].

The BR measurement from ALEPH, DELPHI and other experiments can be compared. This is shown as an example for $B_e$ and $B_{h\pi0}$ in Fig.1. In both case ALEPH has the most precise measurement, which is also true for all those decay modes with a BR greater than 1%, below which CLEO at CESR is more precise due to their larger statistical sample.

SFs for decay modes with net strangeness in the final state have been determined by OPAL [4] covering for 93.4% of these final states (the rest is based on Monte Carlo simulations). The normalization of the SFs from OPAL is however based on world average BRs, which are dominated by the ALEPH measurement [5, 6]. Thus the OPAL and ALEPH strange SFs are correlated.

## 3. Physics implications

Because of its relatively large mass and the simplicity of its decay mechanism, the $\tau$ lepton offers many interesting, and sometimes unique, possibilities for testing and improving the SM. Due to the limited space, in the following I shall briefly mention some of these studies here (details may be found in a recent review in [1]).

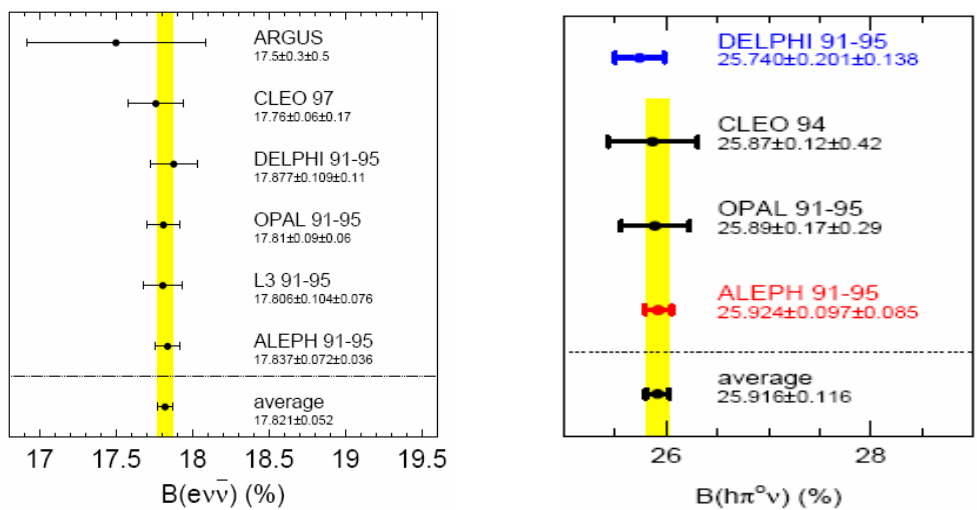

**FIGURE 1.** A comparison of BRs for $e$ leptonic and the largest hadronic decay mode $h\pi^0$, with $h$ standing for both charged $\pi$ and K.

Based on the averaged $e$ and $\mu$ BR measurements, the ratio $B_\mu/B_e$=0.9726±0.0041 agrees well with the predicted value of 0.972565±0.000009 from the standard V-A theory when universality holds. The latter uncertainty arises essentially from the $\tau$ mass determination dominated by the BES result [8]. The ratio is below one because of the $e$, $\mu$ mass difference. Alternatively the measurements yield the ratio of couplings $g_\mu/g_e$=0.9991±0.0033, consistent with unity.

Together with $\tau$ mass and lifetime measurements and $\mu$ decay rate, $\mu$-$\tau$ and $e$-$\tau$ universalities are tested at 0.3% level. The measurement of $B_\pi$ also permits an independent test of $\mu$-$\tau$ universality at 0.5% level.

Since the non-standard $\tau$ decay modes are limited at 0.11% at 95% confidence level [9], one can assume the sum of the leptonic and hadronic BRs is equal to one. Therefore the total $\tau$ hadronic decay rate $R_\tau$ can be obtained from more precise leptonic BRs as $R_\tau$=$B_{had}/B_e$=(1-$B_e$-$B_\mu$)/$B_e$. This quantity can be accurately predicted as function of $\alpha_s(m_\tau^2)$ giving a large perturbative correction of ~21% and a small nonperturbative contribution thereby allowing a precise determination of $\alpha_s$ value at $\tau$ mass scale. The smallness of the nonperturbative contributions are experimentally verified by fitting separately the measured nonstrange vector ($V$) and axial-vector ($A$) part of the SFs and their moments (mass distributions weighted by factor $(1-s/m_\tau^2)^k (s/m_\tau^2)^l$) with the corresponding predictions expressed in terms of $\alpha_s$ and nonperturbative components. Indeed the $V$ and $A$ fits show that the resulting $\alpha_s$ values are consistent whereas the small nonperturbative contributions differ mostly in sign and largely cancel making the latter contributions even smaller in the $V+A$ analysis. It is thus believed that the $\alpha_s$ value $\alpha_s(m_\tau^2)$=0.345±0.004$_{exp}$±0.009$_{th}$ derived in the fit to the $V+A$ distributions is the most robust one. When it is extrapolated to the $Z$ mass, its value $\alpha_s(M_Z^2)$=0.1215±0.0012$_{total}$ is in perfect agreement with the independent measurement from the Z width at LEP, $\alpha_s(M_Z^2)$ =0.1186± 0.0027. This is one of the most precise tests of the running of $\alpha_s$ as predicted by QCD.

The nonstrange $\tau$ vector SFs can be compared with $e^+e^-$ annihilation data via CVC [1]. It is found while the $\tau$ data from different experiments are in good agreement; there is mass-dependent discrepancy between $\tau$ and $e^+e^-$ data (e.g. CMD-2 [10] and KLOE [11]) and also among different $e^+e^-$ data. The new data recently made available by SND [12] are however in better agreement with $\tau$. The prediction derived using both $\tau$ and $e^+e^-$ data (SND data not yet included) on the lowest-order hadronic vacuum polarization component $a_\mu^{had,LO}$, which is the leading contributor to the uncertainty of the muon anomalous magnetic moment g-2, can be compared with the measurement dominated by E821 [13] giving results which are respectively consistent ($\tau$) and at 2.7 standard deviation ($e^+e^-$).

The strange component of the $\tau$ hadronic width gives access to the s quark mass although there is some concern on the stability of the $m_s$ determination [1]. On the other hand, the determination of the CKM matrix element $|V_{us}|$=0.2204±0.0028$_{exp}$±0.0003$_{th}$±0.0001$_{ms}$ using the nonweighted strange SF with maximum inclusiveness and a value $m_s(m_\tau^2)$ =79±8 MeV from lattice calculations [14] appears on safe grounds.

Apart from the BR measurements, the SFs and the resulting determinations can significantly benefit from using larger statistical samples provided by the B Factories. In parallel on theory side when higher order calculations become available further progress in the QCD analyses is expected.

**Acknowledgement**

The author wishes to thank colleagues from LEP experiments for the results shown here and in particular M. Davier and A. Höcker for fruitful collaborations.

**References**

[1] M. Davier, A. Höcker and Z. Zhang, *The physics of hadronic $\tau$ decays*, submitted to *Rev. Mod. Phys.*, [hep-ph/0507078].
[2] ALEPH Collaboration (S. Schael et al.), *Phys. Rept*. **421** (2005) 191.
[3] F. Matorras et al. (DELPHI Collaboration), contributed paper number 094 to the conference.
[4] OPAL Collaboration (G. Abbiendi et al.), *Eur. Phys. J*. C35 (2004) 437.
[5] ALEPH Collaboration (R. Barate et al.), *Eur. Phys. J*. **C1** (1998) 65; **C4**(1998)29; **C10** (1999) 1.
[6] ALEPH Collaboration (R. Barate et al.), *Eur. Phys. J*. **C11** (1999) 599.
[7] DELPHI Collaboration (P. Abreu et al.), *Eur. Phys. J*. **C10** (1999) 201.
[8] BES Collaboration (J.Z. Bai et al.), *Phys. Rev. Lett*. **92** (2002) 101802.
[9] S. Snow, Proceedings of the 2$^{nd}$ international workshop on t lepton physics, Colombus 1992, K.K. Gan ed., World Scientific (1993).
[10] CMD-2 Collaboration (R.R. Akhmetshin et al.), *Phys. Lett*. **B578** (2004) 285.
[11] KLOE Collaboration (A. Aloisio et al.), *Phys. Lett*. **B606** (2005) 12.
[12] SND Collaboration (M.N. Achasov et al.), [hep-ex/0506076].
[13] Muon (g-2) Collaboration (G.W. Bennett et al.), *Phys. Rev. Lett*. **92** (2004) 161802.
[14] C. Aubin et al., *Phys. Rev*. **D70** (2004) 031504.